# Tracking and Quantifying Censorship on a Chinese Microblogging Site

Tao Zhu
zhutao777@gmail.com
Independent Researcher

David Phipps
Computer Science
Bowdoin College

Adam Pridgen
Computer Science
Rice University

Jedidiah R. Crandall
Computer Science
University of New Mexico

Dan S. Wallach
Computer Science
Rice University

*Abstract*—We present measurements and analysis of censorship on Weibo, a popular microblogging site in China. Since we were limited in the rate at which we could download posts, we identified users likely to participate in sensitive topics and recursively followed their social contacts. We also leveraged new natural language processing techniques to pick out trending topics despite the use of neologisms, named entities, and informal language usage in Chinese social media.

We found that Weibo dynamically adapts to the changing interests of its users through multiple layers of filtering. The filtering includes both retroactively searching posts by keyword or repost links to delete them, and rejecting posts as they are posted. The trend of sensitive topics is short-lived, suggesting that the censorship is effective in stopping the "viral" spread of sensitive issues. We also give evidence that sensitive topics in Weibo only scarcely propagate beyond a core of sensitive posters.

## I. INTRODUCTION

Many countries want to have the benefits of the Internet while also controlling the dissemination of content. There are many reasons for censoring the Internet, but a major reason is to prevent civil unrest [1]. Censorship is difficult to maintain at scale [2]. Social media companies in countries that tightly control Internet content find themselves in a risky position. Without lively discussions, they will not have user engagement and the advertising revenues that come with it. If discussions turn to sensitive topics, they risk government intervention that could have negative consequences for the company. Consequently, many Western social network sites, such as Twitter or Facebook, are unavailable inside nations such as China, and homegrown competitors, such as Sina Weibo, have developed sophisticated censorship systems.

Weibo is the most popular microblogging site in China. In February 2012, it had over 300 million users [3]. Like Twitter in other countries, Weibo plays an important role in the discourse surrounding current events in China. Both professional reporters and amateurs can provide immediate, first-hand accounts and opinions of events as they unfold. Even though a user might normally only see messages from other users that they directly follow, any message can rapidly "go viral" by being shared or "retweeted" from one user to the next. For companies such as Twitter or Facebook, viral posts impose technical challenges to the scalability and response time of their service. To Weibo and other such firms, they create huge censorship challenges as well. Weibo must conduct its own censorship rather than relying on external mechanisms, such as China's "Great Firewall."

Since the media is heavily controlled in China, some current events do not appear in traditional media, and are not as prominent in social media as less sensitive topics are. We implemented a system which can find these current events and the related discussions in social media before such content is removed for censorship reasons. We have developed a user tracking method for finding sensitive topic posts before they are deleted. We also apply novel natural language processing methods. To our knowledge, ours is the first system that can successfully extract topics from posts that use unknown words or phrases in Chinese social media. We also document various proactive and reactive filtering and deletion methods applied by Weibo, and find that these mechanisms are effective at restricting discussion of sensitive topics to a core set of posters.

The rest of this paper is structured as follows. Section II gives some basic background information about microblogging and Internet censorship in China. Then Section III describes the algorithms and methods we used for our measurement and analysis, followed by Section IV that describes the natural language processing we applied to the data and presents results from topical analysis. Section V describes the various mechanisms that Weibo uses to filter sensitive content. Then, after discussion in Section VI, we conclude.

## II. BACKGROUND

China has invested heavily in a mixture of Internet filtering technologies [4], [5], [6], [7], cyber police surveillance [8], [9], [10], and policy [11], [12] in order to tightly control access to, and the spread of, Internet content.

Starting from 2010, when microblogs debuted in China, not only have there been many top news stories where the reporting was driven by social media, but social media has also been part of the story itself for a number of prominent events, including the protests of Wukan [13], the Deng Yujiao incident [14], the Yao Jiaxin murder case [15] and the recent Shifang protest [16]. There have also been events where social media has forced the government to address issues directly, such as the Beijing rainstorms in July.

Sina Weibo (`weibo.com`, referred to in this paper simply as "Weibo") has the most active user community of any microblog site in China [17]. Weibo provides services which are similar to Twitter, with @usernames, #hashtags, reposting, and URL shortening. Posts can also contain 140 UTF-8 characters. Because words in written Chinese often only require two characters, 140 Chinese characters can have 3-5 times as much information as 140 characters in English. Two other major differences are that Weibo allows embedded images and videos, and users can comment on each others' posts.

There is relatively little research on Chinese social media [18], despite the fact that English-speaking social media scholars perform research on Russian and Persian language online corpora even though they cannot speak Russian or Persian (see, *e.g.*, [19]). One of many concerns that can hinder this work is the general difficulty of mechanically processing Chinese text. Western speakers (and algorithms) expect words to be separated by whitespace or punctuation. In written Chinese, however, there are no such word boundary delimiters. The word segmentation problem in Chinese is exacerbated by the existence of unknown words such as named entities (people, companies, movies, etc.) or neologisms (substituting characters that appear similar to others, or otherwise coining new euphemisms or expressions, to defeat keyword-based censorship) [20]. Furthermore, since social media is heavily centered around current events, it may well contain new named entities that will not appear in any static lexicon [21].

Despite these concerns, Weibo censorship has been the subject of previous research. Bamman *et al.* [22] performed a statistical analysis of deleted posts, showing that the presence of some sensitive terms indicated a higher probability of the deletion of a post. Their work also showed some geographic patterns in post deletion, with posts from the provinces of Tibet and Qinghai exhibiting a higher deletion rate than other provinces.

WeiboScope [23] was developed at the University of Hong Kong in order to track trends on Weibo, concurrently with our own study of Weibo. The core difference between our work and WeiboScope is that they track a large sample: around 300 thousand users who each have more than 1000 followers, allowing them to observe usually no more than 100 deletions per day. We follow a much smaller sample of users, as we will describe shortly, but we try to focus our attention on users who are more likely to be censored, enabling us to see many more censorship events per day.

## III. POST DELETION DETECTING SYSTEM

Extracting information from Weibo is not all that different from the sort of crawling that must be performed by any search engine. In Weibo each IP address and Application Programming Interface (API) has a rate limit for access the service. This forced us to make a number of engineering compromises, notably focusing our attention where we felt we could find those posts most likely to be subject to censorship.

### A. Identifying the user group

Which users are most likely to be censored? We started with 25 sensitive users that we discovered manually, leveraging a list from China Digital Times [24] of sensitive keywords which are not allowed to be searched on Weibo's server at any given time. To find our initial sample of users who post about sensitive topics, we searched the out-dated keywords that were later allowed to be searched. From the search results, we picked 25 users who stood out for posting about sensitive topics.

Next, we needed to broaden our search to a larger group of users. We assumed that anybody who has been reposted more than five times by our sensitive users must be sensitive as well. We followed them for a period of time and measured how often their posts were deleted. Any user with more than 5 deleted posts was considered to be sensitive. By expanding outward in this fashion, after five days of crawling we settled on a group of over 3,500 sensitive Weibo users, collectively experiencing more than 3,000 deletions per day (roughly 4% of the total posts from those users).

While our methodology cannot be considered to yield a representative sample of Weibo users overall, we believe it is representative of how users who discuss sensitive topics will experience Weibo's censorship. We also believe our sample is broad enough to also speak to the topics that Weibo is censoring on any given day.

### B. Crawling

Once we settled on our list of users to follow, we wanted to follow them with sufficient fidelity to see posts as they were made and measure how long they last prior to being deleted.

We use two APIs provided by Weibo, allowing us to query individual user timelines as well as the public timeline [25]. Starting in July 2012, we queried each of our 3500 users, once per minute, for which Weibo returns the most recent 50 posts. We also queried the public timeline roughly once every four seconds, for which Weibo returns the most recent 200 posts. These posts appear to be 2-5 minutes older than real-time. We believe they represent anywhere from 1.7% to 10% of the full public timeline (*i.e.*, we seem to obtain more complete traces at night, when many of Weibo's users are sleeping and the posting rate is correspondingly lower).

Weibo appears to enforce rate limits both on individual users' queries as well as on source IP addresses, regardless of what user account is being used for the query. To overcome these concerns, we used roughly 300 concurrent Tor circuits [26], driven from our research computing cluster. (We note, for future researchers who may find Tor blocked, that network proxies are easy to come by. Amazon's EC2 and many other companies sell such services.) Our resulting data was stored and processed on a four-node cluster using Hadoop and HBase [27].

### C. Detecting deletions

Here today and gone tomorrow: an absent post may have been censored, or it may have been deleted for any of a variety of other reasons. User accounts can also be closed, possibly for censorship purposes. Users cannot delete their own account, only the system can delete accounts. We conducted a variety of short empirical tests to see if we could distinguish the different cases.

When a post is censored, Weibo will no longer show the post in the public timeline or in a user's individual timeline.

However, it is still there. Using the identifier for the post the post can still be queried, resulting in, "permission denied." Similar behavior occurs when a user's account is deleted. However, when a user directly deletes a post, it appears to be gone and cannot be fetched again by any API call of which we are aware. Instead the message is, "Post does not exist." Consequently, we can now distinguish between these different types of deletions.

Our crawler, which fetches each sensitive user's timeline iteratively, is looking for missing posts. If a post is in our database but is not returned from Weibo, then we check to see if the post is actually still there or, if not, what error message is returned. Ultimately, with the speed of our crawler, we can detect a censorship event within a minute of its occurrence.

## IV. TOPIC EXTRACTION

Even though we are following a relatively modest number of Weibo authors, the volume of text we are capturing is still far too much to process manually. We need automatic methods to classify the posts that we see, particularly the ones which are deleted. For this, we will borrow and extend techniques from the *topic extraction* literature.

Automatic topic extraction is the process of identifying important terms in the text that are representative of the corpus as a whole. Topic extraction was originally proposed by Luhn [28] in 1958. The basic idea is to assign weights to terms and sentences based on their frequency and some other statistical information.

However, when it comes to microblog text, standard language processing tools become inapplicable [29], [30]. Microblogs typically contain short sentences and casual language [31]. Unknown words, such as named entities and neologisms often cause problems with these term-based models. It can be especially challenging to extract topics from Asian languages such as Chinese, Korean, and Japanese, which have no spaces between words. As such, traditional topic extraction techniques based on lexical overlap (use of the same words) become difficult to apply.

We applied the Pointillism approach [18] and TF*IDF to extract hot topics. In the Pointillism model, a corpus is divided into fixed-length grams; words and phrases are reconstructed from grams using external information (*e.g.*, temporal correlations in the appearance of grams), giving the context necessary to manage informal uses of the language such as neologisms. Salton's TF*IDF [32], which stands for "term frequency, inverse document frequency," assigns weights to the terms of a document based on the terms' relative importance to that document compared to the entire corpus. For example, very common words such as the word "the" in English are given a very low weight because the appearance of "the" carries very little information. We next explain how these techniques work together.

### A. Algorithm

TF*IDF is a common method to discover the importance of certain words of a document in a corpus. The TF*IDF value in our case is calculated as:

$$f(t, d_{day}) \times log \frac{\text{Total number of posts for the month}}{f(t, d_{month})}$$

Here, $f(t, d)$ means the frequency of the term $t$ in document $d$. We use trigrams as $t$, and documents $d$ are sets of tweets over a certain period of time. $d_{day}$ is the deleted tweets we caught on day $day$. Similarly, $d_{month}$ is the total deleted tweets we caught in month $month$. IDF is the inverse of the document frequency for deleted posts per month.

First we calculate TF*IDF scores for all trigrams that have more than 20 occurrences in a day. The top 1000 trigrams with the highest TF*IDF score will be fed to our trigram connection algorithm, hereafter "Connector." To connect trigrams back into longer phrases, Connector finds two trigrams which have two overlapping characters. For instance, if there are ABC and BCD, Connector will connect them to become ABCD. Sometimes there is more than one choice for connecting trigrams, *e.g.*, there could also be BCE and BCF. Sometimes trigrams can even form a loop. To solve these problems, we first build directed graphs for the trigrams with a high TF*IDF score. Each node is a trigram, and edges indicate the overlap information between two trigrams. For example, if ABC and BCD can be connected to make ABCD, then there is an edge from 'ABC' to 'BCD'. After all trigrams are selected, we use DFT (Depth First Traversal) to output the nodes. During the DFT we check to see if a node has been traversed already. If so we do not traverse it again. We use the notation '...' to express previously traversed nodes. After the graphs have been traversed, we obtain a set of phrases.

For example, the Connector output of the third most popular topic on 4 August 2012 is:

1.头骨进京鸣冤。河北广平县上坡村76岁的农民冯虎，其子在19

skull go Beijing to redress an injustice. The son of a 76 year old farmer Fenghu, from Shangpo village, Guangping city, Hebei province, was ... at 19

2.头骨进京鸣冤。冯出示的头骨赴京鸣...

skull go Beijing to redress an injustice. The skull shown by Feng go Beijing to redress an injustice...

3.头骨进京鸣冤。冯出示的头骨前额有一大窟窿，他...

skull go Beijing to redress an injustice. There is a big hole on the skull shown by Feng, he...

4.头骨进京鸣冤。冯出示的头骨前额有一个无罪的公民...

skull go Beijing to redress an injustice. There is a innocent citizen on the skull shown by Feng, he...

5.头骨进京鸣冤。冯出示的头骨进...

skull go Beijing to redress an injustice. The skull shown by Feng enter...

6.头骨进京鸣冤。冯出示的头等舱

skull go Beijing to redress an injustice. The first class seat shown by Feng...

7.【華聯社電】上访15年 老父携儿头骨...

Chinese Community report: petition 15 years, old father bring the skull of his son...

Outputs 4 and 6 are incorrectly connected. This is because the same trigrams are shared by different stories that have high TF*IDF scores on the same day. This problem can be solved by examining the cosine similarity of the frequency of occurrence of the first and the last trigram for each result.

Cosine similarity is used to judge whether two trigrams have correlated trends.

$$cos.Sim = \frac{<A_i, B_i>}{\sqrt{\sum_{i=1}^{n} A_i^2} \times \sqrt{\sum_{i=1}^{n} B_i^2}}$$

where $<,>$ denotes an inner product between two vectors. For details, please refer to Song *et al.* [18].

### B. Results

From the connected sentences we can begin to understand the general events that are driving major sensitive topics of discussion on Weibo. Table I and Table II list the top 5 topics of the deleted posts from 20 July 2012 to 20 August 2012. (A computer failure in our cluster prevented us from collecting data on 6 August 2012.) Note that we just translated the posts from each topical cluster, we have not confirmed the veracity of any of the claims of the Weibo users' posts that we translated.

Interestingly, besides named entities, we also extracted three neologisms. They are 李W阳 (Li Wangyang, from 李旺阳), 六圖四 (June Fourth, from 六四), and both 启-东 and 启/东 (Qidong, from 启东).

Next, we estimate how much deleted posts affect a topic's propagation by demonstrating the frequency changes by day. By using the proposed technique from Section IV-A, we extract the top deleted topics, with a trigram representing that topic in the sensitive user group.

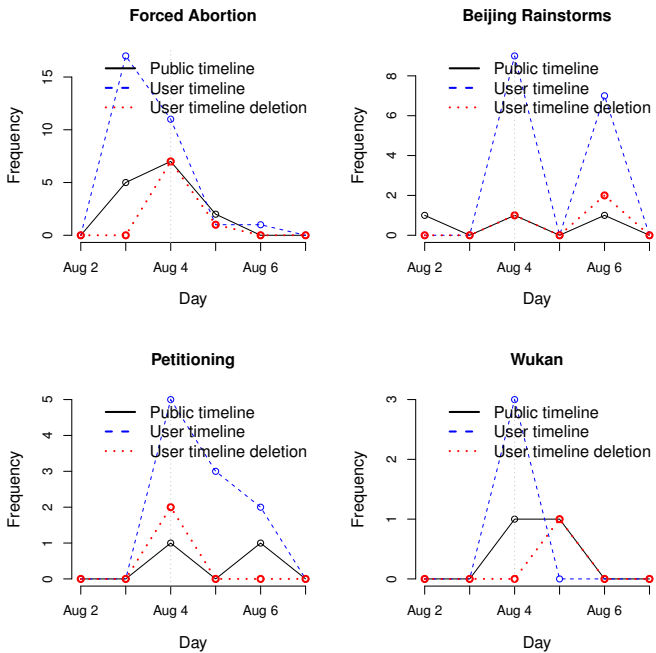

Fig. 1. **Deletion efficiency in the short term.**

Figure 1 shows the trends of the top four censored topics on 4 August 2012. Each sub-graph shows the trigram's frequency changes among three different sets of data that we have: the blue dashed line shows the trigram's frequency in the user timeline, the red dotted lines tell us how many times the trigram were deleted from the user timeline, and the black solid line shows how many times that trigram appeared in the public timeline. Note that we do not include reposts in this graph, since the public timeline does not include any reposts.

The first topic is about a forced abortion accident that happened three years ago in which both the mother and the unborn baby died. We see the topic's frequency was 0 in both datasets on 2 August 2012 and peaked on 3 August 2012 in the user timeline when it was reported. On 4 August 2012, the number of occurrences of this term in the public timeline started to increase (from 5 to 7), but as censorship started, the frequency in the user timeline began to drop, and it disappeared entirely on 7 August 2012.

Our public timeline has around 17 times more data than our user timeline dataset, so usually trigrams in the public timeline will have a larger frequency than in the user timeline. However, in this example and the three others we see the opposite. This means that the topic did not spread into the public timeline, which could be considered successful in terms of achieving a censors' goal of preventing the "viral" spread of a particular topic.

In the other three examples we observed similar results. Starting from the point where censorship begins, the frequency of the topic drops dramatically. This prevents users from discussing the topic with like-minded individuals or getting feedback, causing them to lose interest and stop posting the sensitive content. We begin to see how effective Weibo's censorship campaigns can be. Despite starting well after a post has begun to circulate, Weibo can effectively prevent discussion of sensitive content from going "viral" across its site.

## V. WEIBO CENSORSHIP MECHANISMS

Sina Weibo has a complex variety of censorship mechanisms, including both proactive and retroactive mechanisms.

Proactive mechanisms may include:

- **Explicit filtering:** Weibo will inform a poster that their post cannot be released because of sensitive content, with a message such as, "Sorry, since this content violates 'Sina Weibo regulation rules' or related regulation or policy, this operation can't be processed. If you need help, please contact customer service."
- **Implicit filtering:** Weibo sometimes suspends posts until they can be manually checked, with a message such as, "Your post has been submitted successfully. Currently, there is a delay caused by server data synchronization. Please wait for 1 to 2 minutes. Thank you very much."
- **Camouflaged posts:** Weibo also makes it appear to a user that their post was released, but other users are not able to see the post. The poster receives no warning message in this case. One example we observed is on 1 August

TABLE I
Blocked Topics - part 1.

| | Top 1 | Top 2 | Top 3 | Top 4 | Top 5 |
|---|---|---|---|---|---|
| 7-20 | Syria rebels banner[1] | Support Syria rebels | | | |
| 7-21 | Ji County[2] | Syria rebels banner[1] | Liu Futang[3] | Support Syria rebels | |
| 7-22 | Beijing rainstorms[4] | Subway collapse[5] | Liu Futang[3] | Guo Jinlong[6] | |
| 7-23 | Fanshan Geracomium[7] | Guo Jinlong[6] | Beijing rainstorm toll[8] | | |
| 7-24 | Donate your sister[9] | Inspect[10] | | | |
| 7-25 | Beijing rainstorm toll[8] | Guo Jinlong[6] | Qidong[11] | Taiwan[12] | |
| 7-27 | Red Cross scandal[13] | Taiwan[14] | Qidong[11] | Fangshan[15] | |
| 7-28 | Qidong[11] | Liu Futang[3] | | | |
| 7-29 | Japanese reporter[16] | Gu Kailai's case[17] | Protest in Hong Kong[18] | Beijing rainstorms[4] | |
| 7-30 | Zhou Jun[19] | Political reforms[20] | Protest in Hong Kong[18] | Establishment[21] | 李W阳[22] |
| 7-31 | Yao Wenwei[23] | Protest in Hong Kong[18] | Qidong[11] | Accident63 | Complain gov.[24] |
| 8-1 | Complain gov.[24] | General Liu Yazhou | Gu Kailai's case[17] | | |
| 8-2 | Freedom of speech[25] | Red Cross scandal[26] | General Liu Yazhou | Edmund Valtman | Demolish old buildings[27] |
| 8-3 | One Child Policy abuse[28] | Despise CPC[29] | Cultural Revolution[30] | Chairman Mao[31] | Beijing rainstorms[4] |
| 8-4 | One-Child Policy abuse[28] | Beijing rainstorms[32] | Human Rights News | Wukan[33] | Protect environment[34] |
| 8-5 | Human Rights News[35] | Freedom of speech | A social phenomena[36] | Qidong (启-东)[37] | Gu Kailai's case[17] |
| 8-7 | Qian Yunhui Accident[38] | Despise gov. | Gov. scandal | Gov. scandal | Despise gov. |
| 8-8 | Qian Yunhui accident[38] | Gov. scandal[39] | Beijing rainstorms | Group sex[40] | Liu Futang[3] |
| 8-9 | Group sex[40] | Li Wangyang[41] | Gu Kailai's case[17] | Liu Futang[3] | Gov. scandal |

| | |
|---|---|
| 1 | "Syria proves it's a proxy war against Russia and China." The banner says, "Russia and China, whatever you do to protect your puppet will not stop us from kicking your asses out of Syria." Picture available at http://twitpic.com/9u9bjz |
| 2 | People questioning the Ji County (in Tianjin) fire death toll. |
| 3 | An environmentalist who was arrested on July 20th from a hospital. http://www.chinadialogue.net/article/show/single/en/4866 |
| 4 | http://www.businessweek.com/articles/2012-07-23/giant-beijing-rainstorm-triggers-citizens-anger |
| 5 | Rumor about a subway collapse in Beijing. |
| 6 | Former mayor of Beijing. Although he resigned as mayor of Beijing, it was said that it was not a resignation in response to the Beijing rainstorms. |
| 7 | A rumor saying that 200 old people were hit by the Beijing rainstorms. |
| 8 | Posts calling into question the Beijing rainstorm death toll. |
| 9 | Refusal to donate for Beijing rainstorm relief efforts because some people believe that the government did not perform well in providing disaster relief. |
| 10 | Perception that the leaders did not perform well when inspecting the condition of the rainstorm disaster. |
| 11 | Protest against waste project. http://www.reuters.com/article/2012/07/28/us-china-environment-protest-idUSBRE86R02Y20120728 |
| 12 | Taiwanese flag pulled by Olympics organizers after a complaint by Chinese officials. |
| 13 | The Red Cross rejected the family of a victim in the Beijing rainstorms trying to get back the body and trying to "rescue" the body which had been in water for two days. They charged the family for the rescue fee. |
| 14 | ROC will be at half-mast to mourn the 77 victims of the Beijing rainstorms. |
| 15 | There are news posts with pictures showing the police Commissioners helped to salvage corpses. But net users said they were just putting on a show. |
| 16 | A Japanese reporter was beaten by police when he was interviewing a participant in the Qidong protest. |
| 17 | http://en.wikipedia.org/wiki/Gu_Kailai |
| 18 | Protest against National Education in Hong Kong. http://www.youtube.com/watch?v=qYyJqGfyEDc |
| 19 | Net users criticizing the mass media's impropriety on Zhou Jun's failure in the Olympic Games. |
| 20 | Wen Jiabao's words about political reforms in a CNN interview. |
| 21 | The phrase "Establishment hijack the country." |
| 22 | Neologism for 李旺阳 (Li Wangyang). http://en.wikipedia.org/wiki/Death_of_Li_Wangyang |
| 23 | A former official was sentenced for inciting subversion of state power because he wrote 52 articles to appeal judicial independence. |
| 24 | Complaint about the government for discriminating against farmers. |
| 25 | Supreme court judge ordered Weibo to give Ms. Yu 2520 Yan for closing her Weibo account twice. |
| 26 | The Red Cross asked for 720 yen to deliver the dead bodies during the Beijing rainstorms, regardless if the person has died or the family wanted to deliver the body by themselves. |
| 27 | An accident caused by demolishing old buildings. |
| 28 | A forced abortion accident that happened three years ago which caused both the mother and the unborn baby to die. |
| 29 | Mao's kaiserism. |
| 30 | A tragedy that happened in 1970 during the Cultural Revolution. |
| 31 | Chairman Mao Memorial Hall is in application for a World Heritage site. |
| 32 | Complaints that the government did not update the death toll number caused by the Beijing rainstorms. |
| 33 | In the farmers' representative meeting, Wukan village decided to turn over the wall of the land sold by former officials. |
| 34 | Appealing to help Taizhou, it will become Shifang soon. http://en.wikipedia.org/wiki/Shifang |
| 35 | Li Guizhi, a civil rights activist, was arrested again after escaping from police. |
| 36 | College graduates are eager to become officials of the government. |
| 37 | A perception that Qidong started to arrest people wildly. |
| 38 | New evidence regarding Qian Yunhui's death. http://en.wikipedia.org/wiki/Qian_Yunhui |
| 39 | Some corruption stories related to the former deputy mayor of Maoming city, Guangdong province. |
| 40 | There are a series of indecent photos which contain six people (three female and 3 male) having group sex. Net users tried to identify these six people and found that they are from Lujiang and some of them are officials. |
| 41 | Zhu Zhicheng was arrested, allegedly for not cooperating with the government to hide the reason why Li Wangyang died. |

TABLE II
Blocked Topics - part 2.

| | Top 1 | Top 2 | Top 3 | Top 4 | Top 5 |
|---|---|---|---|---|---|
| 8-10 | RTL | Freedom of speech[42] | Group sex[40] | Gu Kailai's case[17] | |
| 8-11 | Group sex[40] | Gu Kailai's case[17] | DHBTC[43] | Police scandal[44] | June Fourth (六圖四)[45] |
| 8-12 | Tang Hui[46] | Group sex[40] | Kong Qingdong[47] | RTL[59] | DHBTC[43] |
| 8-13 | DHBTC[43] | RTL[59] | Despise gov.[48] | Gu Kailai's case[17] | |
| 8-14 | Hongkong[49] | Scandal of gov.[50] | Tang Hui[51] | Zhou Kehua[52] | DHBTC[43] |
| 8-15 | DHBTC[43] | | | | |
| 8-16 | Rumor[53] | Zhou Kehua[52] | Diaoyu Island[54] | | |
| 8-17 | Anti-Japanese[55] | DHBTC[43] | North Korea[56] | Scandal of gov.[57] | Cultural Revolution[58] |
| 8-18 | Zhou Kehua[52] | Anti-Japanese[62] | DHBTC[43] | | |
| 8-19 | Anti-Japanese[62] | RTL[59] | Zhou Kehua[52] | | |
| 8-20 | Anti-Japanese[62] | Freedom speech[60] | Zhou Kehua[52] | Gu Kailai[61] | |

| | |
|---|---|
| 42 | Ren Jianyu was sentenced to re-education through labor (RTL) for two years for reposting others' posts. |
| 43 | The bodies and parts currently on display in New York are licensed to the Premier by the Dalian Hoffen Bio Technique Company Ltd. (DHBTC). DHBTC acquired the bodies indirectly from the Chinese Bureau of Police, which deemed them unclaimed at death. |
| 44 | Li Hongna posted a story about herself saying that she was raped by 8 people, including the police inspector. |
| 45 | The Tiananmen Square protests of 1989, or June Fourth Incident. |
| 46 | Tanghui was sentenced to 2 years of RTL for petitioning for the case of her daughter. When her daughter was 11 years old, she was forced to become a prostitute. http://zh.wikipedia.org/zh/唐慧劳教案 |
| 47 | Kong Qingdong says: "Taxpayers, to go the hell." http://en.wikipedia.org/wiki/Kong_Qingdong |
| 48 | A work of fiction about government corruption. |
| 49 | Hongkong college students did not welcome astronauts visiting from China. |
| 50 | Municipal administration beat a passerby who took pictures while they were beating a peddler. After that the municipal administration said their privacy had been invaded. |
| 51 | The daughter of Tanghui has allegedly been raped by government officers. |
| 52 | People are suspicious that the killed person is not Zhou Kehua. |
| 53 | This is a rumor: Li Ning, a university student, has been made into an exhibition body by Dalian Hoffen Bio Technique Company Ltd. Li Ning's mother died suspiciously in the police office. Li Ning petitioned several times in Beijing. |
| 54 | Argument about who owns Diaoyu Island, China or Japan. |
| 55 | Hongkong people complained the Chinese government are cowards to let the Japanese occupy Diaoyu Island. |
| 56 | In response to the rumor that the Chinese government agreed to loan money to North Korea, people saying they do not agree with it. |
| 57 | An original government officer, Zhou Wenbin, confessed that he had offered bribes to Yuwei, an officer at the Land Resources Bureau of Lixin, Anhui province in 2011. Zhou Wenbin started a hunger strike starting from the next day until the personnel concerning the case are arrested. |
| 58 | Fu Dongmei, the daughter of Fu Zuoyi, was treated badly during the Cultural Revolution. |
| 59 | Wen Jiabao said that RTL should be prohibited. |
| 60 | A user's account had been blocked, after it was unblocked, he appealed to all the users whose accounts had been blocked to become a union. He asked several questions: How to protect the freedom of speech?; What kind of organization is the State Internet Information Office? ; and so on. |
| 61 | People do not agree with the suspended death sentence given to Gu Kailai. They called Bo Xilai, Gu Kailai, and family the biggest mafia in China. |
| 62 | Statements that the "Angry Youth" who damaged the Japanese cars on the street are actually traitors. |
| 63 | A sinkhole happened in Changan street, Beijing. |

2012 for a post containing the substring 'cgc' (Cheng Guangcheng [33]).

- **Surveillance:** While we have no evidence specific to Weibo, it is worth pointing out here that surveillance is often coupled with censorship [10].

When Weibo discovers that posts containing sensitive content have passed through the filters described above, Weibo will stop the "viral" propagation of such content. In May 2012, Sina Weibo announced a "user credit" points system [34] through which users can report sensitive or rumor-based posts to the administrators. However, while the points system is used to stymie the spread of rumors through the Weibo community, removal of sensitive content is still carried out without the transparency of a points system.

Retroactive mechanisms for removing content that has already been released may include:

- **Backwards reposts search:** In our deleted posts dataset over 82% of reposted posts have a standard deviation of less than 5 minutes deletion time for the ones repost the same post, meaning that a sensitive post is detected and then most of the other posts in a chain of reposts are then deleted.

- **Backwards keyword search:** We also observed that Weibo sometimes removes posts retroactively in a way that causes spikes in the deletion rate of a particular keyword within a short amount of time. For example, there is a neologism 夭朝 (Tian Chao, an offensive name for China, originally from 天朝, or Tian Chao, same pronunciation as 夭朝, which means Celestial Empire and is a derogatory term for today's China because it implies that it is similar to the feudalistic past. 夭朝 makes the connotation worse, because 夭 contains the two characters 王八, meaning "bastard.". The frequency of occurrence of 夭朝 in deleted posts is the sequence (2,0,1,6,3,0,0,2,2,0,3,0,2,3,3,2,1,2,0,0,1,0,0,0,5,4,4,2,**14**,3,6,4)

respectively from 25 July 2012 to 25 August 2012. There is a concentrated deletion (14) of this word on 22 August 2012. We examined each deleted post on 22 August for this keyword and confirmed that these deletions were not caused by other mechanisms that we know about.

- **Monitoring specific users:** For this research, we focused our monitoring on users who we believed were more likely to be censored based on their system deletion history. Does Weibo do the same thing? Figure 2 plots this effect. We grouped users together who had the same number of censorship events occur to their posts. The $x$-axis is the number of such deletions. The $y$-axis shows how long these to-be-censored posts live before they are deleted for all the users in each cohort. The black circles show the *median* lifetime of posts in the cohort, and the dotted blue bars show plus or minus one standard deviation from the *mean*. The clear downward trend in the medians can be interpreted in two ways. Either Weibo pays more attention to users once they have been censored, or users who are more profligate with sensitive posts are less inclined to try to work around Weibo's censorship.
- **Account closures:** Weibo also closes users' accounts. There were over 300 user accounts closed by the system from our sensitive user group (out of over 3,500 users).
- **Search filtering:** To prevent users from finding sensitive information on weibo.com, Weibo also has a frequently updated list of words [24] which cannot be searched.
- **Public timeline filtering:** We believe that sensitive topics are filtered out of the public timeline. We do not believe that this affects our results from Section IV where we compared the user timeline to the public timeline; however, because the filtering appears to be limited to only general topics that have been known to be sensitive for a relative long time.

## VI. DISCUSSION

Our results demonstrate that if the goal of censorship is to stop sensitive topics from being discussed for a long period of time and prevent the discussion from spreading throughout the system, then Weibo's censorship mechanisms are effective. Our topical analysis shows that topics usually last just two or three days at the most, and we further showed that topics do not spread from the user timeline (*i.e.*, the users who regularly discuss sensitive topics) to the public timeline (*i.e.*, the majority of users).

An analogy can be drawn with the human immune system. Because users can use neologisms to circumvent proactive filtering of sensitive keywords, Weibo must adapt in much the same way that our body adapts to ever-changing viral threats. Our immune system combines proactive defenses with reactive defenses, and is distributed with multiple layers. Weibo has several layers of filtering with graduated responses, and the ability to retroactively delete posts based on repost linkages and keywords. The detection and removal of sensitive content is also a distributed process. Furthermore, users who often post sensitive content seldom have their accounts removed. They may be an important part of the adaptive immunity process. As Eaves [35] mentioned, Weibo may control the direction of the flow of posts and topics, rather than shutting down the flow.

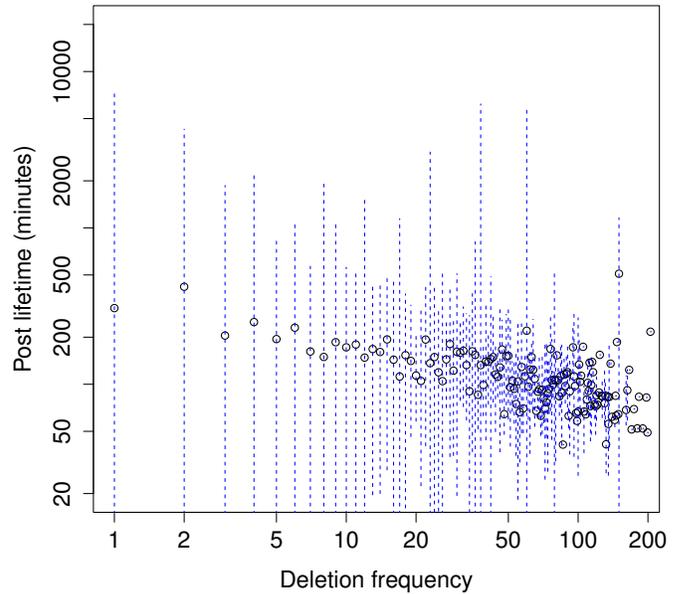

Fig. 2. **Users' median post lifetime in minutes *vs.* the number of deletions for that user on a log-log scale.**

A biological immune system's focus is not on reaching zero infection as quickly as possible, but rather on stopping infections from spreading and interfering with the health of the organism. Weibo's censorship seems to have a similarly pragmatic focus. As long as sensitive topics are not going "viral," a few sensitive posts can be tolerated for some period of time.

Our research detected that deletions happen most heavily in the first hour after a post has been made. Especially for original posts that are not reposts, most deletions occur within 5-30 minutes, accounting for 25% of the total deletions of such posts. Nearly 90% of the deletions of such posts happen within the first 24 hours of the post.

One aspect of censorship that is not considered in our analysis, but would be an interesting topic for future work, is the interactions between social media and traditional media. Leskovec *et al.* [36] gives an interesting analysis of the interplay between blogs and traditional media during the 2008 U.S. Presidential election. Traditional media relevant to Weibo may include the state-run media that is heavily censored, or off-shore news outlets that are uncensored but limited in availability and sometimes offset from China's news cycles by timezone differences.

## VII. CONCLUSIONS

We have shown that Weibo's internal censorship is effective at limiting the amount of time that a sensitive topic is discussed, and at preventing the discussion from spreading from a core set of users who discuss sensitive topics to the general Weibo community. This is accomplished through a combination of proactive and reactive mechanisms. The apparent effectiveness of Weibo's censorship creates difficult technical challenges for building anti-censorship systems. Likewise, other social networking systems may be compelled by other countries to follow Weibo's example, creating public policy challenges for free speech advocates.